\documentclass[sigconf]{acmart}

\AtBeginDocument{%
  }

\setcopyright{acmlicensed}
\copyrightyear{2025}
\acmYear{2025}
\acmDOI{10.1145/3731599.3767578}

\acmConference[SC-W'25]{Workshops of The International Conference on High Performance Computing, Network, Storage, and Analysis}{November 16-21, 2025}{St. Louis, MO}

\usepackage{framed,acmart-taps}
\usepackage{array}
  \usepackage{graphicx}
\usepackage[tight,footnotesize]{subfigure}

\usepackage{listings}
\usepackage{amsfonts}

\usepackage{framed}
\usepackage{xcolor}

\colorlet{shadecolor}{cyan!20}

\newenvironment{myenv}[1]{%
  \MakeFramed{\advance\hsize-\width \FrameRestore}%
  \begin{center}\textbf{#1}\end{center}\vspace{-0.5ex}%
}{%
  \endMakeFramed
}

\RequirePackage{color}
\usepackage{colortbl}
\definecolor{RED}{rgb}{1,0,0}
\definecolor{BLUE}{rgb}{0,0,1}
\definecolor{GREEN}{rgb}{0,1,0}
\definecolor{DGREEN}{RGB}{0,140,51}
\definecolor{color1}{rgb}{0.913, 0.776, 0.686}
\definecolor{color2}{rgb}{0.913, 0.867, 0.686}
\definecolor{ltgray}{rgb}{0.85, 0.85, 0.85}

\usepackage{stackengine}

\newcommand{\modif}[1]{\textcolor{black}{#1}} 
\begin{document}

\title{Do Large Language Models Speak Scientific Workflows?}
\author{Orcun Yildiz}
\email{oyildiz@anl.gov}
\affiliation{%
  \institution{Argonne National Laboratory}
  \streetaddress{9700 S Cass Ave}
  \city{Lemont}
  \state{IL}
  \country{USA}
  \postcode{60439}
}


\author{Tom Peterka}
\email{tpeterka@mcs.anl.gov}
\affiliation{%
  \institution{Argonne National Laboratory}
  \streetaddress{9700 S Cass Ave}
  \city{Lemont}
  \state{IL}
  \country{USA}
  \postcode{60439}
}

 \renewcommand{\shortauthors}{Yildiz et al.}


\begin{abstract} 
In this work, we conduct an experimental study to explore applicability of LLMs for configuring, annotating, and translating scientific workflows. We use three different workflow-specific experiments and evaluate several open- and closed-source language models using state-of-the-art workflow systems. Our studies reveal that LLMs often struggle due to a lack of training data for scientific workflows. We further observe that the performance of LLMs varies across experiments and workflow systems. We discuss the implications of our findings and draw attention to several approaches extending LLM capabilities for scientific workflows. Our findings can help workflow developers and users in understanding LLM capabilities in scientific workflows, and motivate further research applying LLMs to workflows.
\end{abstract}






\maketitle

\section{Introduction}\label{sec:Introduction}



In recent years, we have witnessed the emergence of large language models (LLMs) increasingly being used in many fields beyond text processing. However, current LLMs are not yet tailored for the specialized demands of some scientific tasks and may struggle to perform effectively in these areas. One particularly challenging area for LLMs is scientific workflows, which are often complex, involving large numbers of connected tasks with wide-ranging data and computing requirements.

Over the years, the high-performance computing (HPC) community has developed workflows to support scientists in managing scientific tasks on HPC systems, meeting the demands of today’s increasingly complex and large-scale computing environments. By enabling scientists to define collections of tasks along with their dependencies and data requirements, these workflows automate scheduling and data exchange, simplifying orchestration and potentially accelerating scientific discovery. 

Despite their potential advantages, many scientists find workflow systems difficult to use, often opting to run tasks manually or develop their own workflow solutions instead~\cite{yildiz2024extreme}. Even when generic workflow systems are adopted, scientists often lack understanding of these systems. LLMs could assist in addressing these challenges, but this requires understanding of LLMs' capabilities when applied to HPC workflows.


Several recent studies have explored the application of LLMs to specific HPC-related tasks, from code generation~\cite{nichols2024can}, annotation~\cite{nichols2024hpc}, and  answering queries~\cite{chen2023lm4hpc} to scientific assistants~\cite{mallick2024chatvis, kumar2023mycrunchgpt}. While these works provide important insights into the potential of LLMs for HPC, comprehensive studies on their broader applicability within full workflow systems are still lacking.

In this work, we conduct an experimental study to explore the applicability of LLMs to HPC workflows. We evaluate three distinct workflow-specific tasks using a variety of open- and closed-source language models applied to state-of-the-art workflow systems. 
Our goal is to benchmark LLM performance using parameter settings commonly adopted for HPC-related tasks and to investigate the potential for prompt engineering; we do not explore fine-tuning in this paper. The purpose of this study is to demonstrate the potential for LLM-based agentic systems in scientific workflows and to establish a baseline for future comparisons. We discuss the implications of our findings and highlight several approaches for improving LLM capabilities. We believe that the insights and lessons learned from our experiments will provide a better understanding of LLM strengths and limitations, and motivate further research applying LLMs to scientific workflows.



The remainder of this paper is organized as follows. Section 2 presents background and related work. Section 3 explains our methodology in assessing the capabilities of LLMs for scientific workflows, including a description of the LLMs, workflow systems, experiments, and evaluation metrics. Section 4 presents the results when applying LLMs to a series of different workflow experiments, highlighting the models' strengths and weaknesses.  Section 5 concludes with a summary and brief look at future work.

\section{Background and Related Work}\label{sec:Background}

We first provide a brief background on scientific workflows and large language models. Then, we present the related work on the application of LLMs to workflows.

\subsection{Scientific workflows}


Scientific computing often relies on complex workflows, which comprise a series of tasks that cooperate in terms of scheduling and communication as part of a larger scientific campaign. There are two main types of scientific workflows that we consider in this work: distributed and in situ. Distributed workflows run across several independent systems in a wide area such as grids and clouds, where the tasks exchange data through files. Examples include studies by Deelman et al.~\cite{deelman2019evolution} and Altintas et al.~\cite{altintas2019toward}, along with several surveys summarizing related work~\cite{deelman2009workflows, yu2005taxonomy}. In situ workflows, on the other hand, are designed to run within a single HPC system, with tasks executing concurrently and exchanging data through memory or the HPC system’s interconnect, rather than physical storage. Representatives of such systems include ADIOS~\cite{boyuka2014transparent}, Damaris~\cite{dorier2016damaris}, Decaf~\cite{yildiz2022decaf}, Deisa~\cite{gueroudji2021deisa}, Henson~\cite{morozov2016master}, ParaView Catalyst~\cite{ayachit2015paraview}, SENSEI~\cite{ayachit2016sensei}, VisIt Libsim~\cite{kuhlen2011parallel}, and Wilkins~\cite{yildiz2024wilkins}. 

\subsection{Large language models}

The introduction of transformer-based models~\cite{vaswani2017attention} revolutionized the design of neural networks for understanding and generating human language. Transformer models use a self-attention mechanism to capture the dependencies in data, which allows processing data in parallel instead of sequentially, improving efficiency. Large language models are trained on vast datasets from the web and contain billions of parameters. This in turn enables them to perform well in tasks such as summarization, text classification, content generation, and translation. Examples include GPT-4~\cite{achiam2023gpt}, o3, Claude-4, LLaMA~\cite{touvron2023llama}, Gemini, LaMDA~\cite{thoppilan2022lamda}, and PALM-2~\cite{anil2023palm}.

\subsection{Large language models for scientific workflows}
Duque et al.~\cite{duque2023leveraging} explore using LLMs to build and execute workflows. In particular, they integrate Open AI's function calling API with the Parsl workflow~\cite{babuji2019parsl} system in order to generate workflow functions based on the user's description in natural language. Although the presented framework provides a prototype of a workflow
engine that can execute a variety of tasks, it does not fully explore the challenges associated with applying LLMs to scientific workflows.


Sanger et al.~\cite{sanger2023large} investigate the applicability of GPT-3.5 to three different science use cases for understanding, modifying, and extending scientific workflows. Their findings indicate that while GPT-3.5 performs well in explaining workflows, it struggles with modifying or extending workflows. Although there is some overlap with our study, their work uses GPT-3.5, which has since been followed by improved models (e.g., o3, Gemini-2.5-Pro). \modif{These newer models offer larger context windows and enhanced reasoning capabilities} that we evaluate in this study \modif{to see whether they can address such workflow-specific tasks}. We also provide a broader range of workflow systems and scientific tasks in our study.


\section{Methodology}\label{sec:Methodology}

In this section, we describe our
methodology by presenting descriptions of the LLMs, workflow systems, experiments, and evaluation metrics used in our study.

\subsection{LLMs used for evaluation}

In our experiments, we utilize several state-of-the-art large language models (LLMs), both open-source and closed-source. We selected these models because they are among the top-performing LLMs available. 


\begin{itemize}


\item \textbf{o3} is a closed-source language model from OpenAI. o3 is trained using large-scale reinforcement learning applied to chains of thought providing strong reasoning capabilities~\cite{jaech2024openai}.

\item \textbf{Claude-Sonnet-4} is a closed-source language model developed by Anthropic and designed as a hybrid reasoning model for high-level problem-solving and coding tasks.


\item \textbf{Gemini-2.5-Pro} is Google's most advanced model with strong reasoning and coding capabilities.

\item \textbf{LLaMA-3.3-70B-Instruct}~\cite{touvron2023llama} is an open-source language model from Meta with 70 billion parameters. LLaMA-3.3-70B targets enhancing natural language understanding and generation compared with former LLaMA models.



\end{itemize}
\subsection{Workflow systems}

We use five different state-of-the-art workflow systems to evaluate the LLMs. 

\begin{itemize}

\item \textbf{ADIOS2}~\cite{boyuka2014transparent} is a middleware originally designed as a flexible I/O library for scientific codes. It has been extended to couple simulations with analysis and visualization applications through its I/O interface, facilitating seamless data exchange between different components of scientific workflows.

\item \textbf{Henson}~\cite{morozov2016master} is a cooperative multitasking system for in situ processing. It incorporates a built-in scripting language for workflow description and relies on shared objects and coroutines as its core abstractions. This design allows colocated tasks on the same compute node to exchange data efficiently by passing pointers directly, thereby avoiding data copies.

\item \textbf{Parsl}~\cite{babuji2019parsl} is a Python parallel programming library that enables task-based execution. In its programming model, users define Parsl apps by annotating standard Python functions. When called in a regular Python program, these apps are intercepted by Parsl and sent to the Parsl runtime for execution. Parsl is designed for scalability, offering a flexible set of executors optimized for different use cases such as low-latency tasks, high-throughput workloads, or extreme-scale computing. 

\item \textbf{PyCOMPSs}~\cite{tejedor2017pycompss} is a task-based programming model that enables the development of parallel applications and workflows. It provides task-level parallelization through the annotation of Python methods and a small API. PyCOMPSs is capable of executing on a range of distributed computing platforms, such as clusters, cloud environments, container-managed clusters, and, more recently, edge-to-cloud configurations.

\item \textbf{Wilkins}~\cite{yildiz2024wilkins} is an in situ workflow system that facilitates dynamic heterogeneous task specification and execution for in situ processing. It offers a data-centric API for defining the workflow graph, creating and launching tasks, and establishing communication links between the tasks.



\end{itemize}

\subsection{Description of experiments}

We use three different workflow-specific experiments to evaluate the LLMs.

\paragraph{Workflow configuration} 

Traditionally performed manually by users prior to workflow execution, users define their workflow by indicating data and resource requirements of individual tasks using well-known languages such as YAML or Python. The workflow configuration parameter space is large and high-dimensional, making configuring workflows cumbersome; hence, automating this process through LLMs can help increase productivity of workflow users.

In this experiment, we study the capability of LLMs to generate workflow configuration scripts based on natural language inputs. A sample user prompt is shown below. We evaluate the LLMs based on the correctness of the generated configuration scripts for each workflow system.

\aptLtoX{\begin{shaded}
\begin{center}\textbf{User prompt}\end{center}\par
I would like to have a 3-node workflow consisting of one producer and two consumer tasks, where producer generates grid and particles datasets, consumer1 reads grid and consumer2 reads particles datasets. Producer requires 3 processes, and each consumer runs on a single process.
Please provide the workflow configuration file for the Wilkins workflow system.
\end{shaded}\vspace*{40pt}}{
\begin{myenv}{User prompt}
I would like to have a 3-node workflow consisting of one producer and two consumer tasks, where producer generates grid and particles datasets, consumer1 reads grid and consumer2 reads particles datasets. Producer requires 3 processes, and each consumer runs on a single process.
Please provide the workflow configuration file for the Wilkins workflow system.
\end{myenv}}

\paragraph{Task code annotation} 

Today's workflow systems often require changes to the user task codes when transitioning from manually running these tasks to using a workflow system. Extensive code changes can make the learning curve steeper and create a barrier to entry to using workflows. 

In this experiment, we assess the ability of LLMs to automatically annotate user task codes by asking LLMs to annotate the provided user codes for different workflow systems. Then, we evaluate the correctness of the annotated task codes. A sample user prompt is shown below.

\aptLtoX{\begin{shaded}
\begin{center}\textbf{User prompt}\end{center}\par
You are assisting in the development of a simple producer-consumer workflow using the ADIOS2 system. The producer task code is provided below. Annotate this task code in order to use it with the ADIOS2 system.
\end{shaded}\vspace*{40pt}}
{\begin{myenv}{User prompt}
You are assisting in the development of a simple producer-consumer workflow using the ADIOS2 system. The producer task code is provided below. Annotate this task code in order to use it with the ADIOS2 system.
\end{myenv}}


\paragraph{Task code translation} 


Several related works explore the applicability of LLMs for code translation between different programming languages or execution models. Similarly, our objective is to assess the applicability of LLMs for annotated task code translation between different workflow systems. For instance, users may want to compare different workflow systems for the same problem. Ideally, users would use LLMs for code translation, freeing them from having to learn a new workflow system. A sample prompt is shown below, which evaluates the capability of LLMs to perform task code translation between different workflow systems.

\vspace{0.1in}

\aptLtoX{\begin{shaded}
\begin{center}\textbf{User prompt}\end{center}\par
Task codes are provided below for the PyCOMPSs workflow system for a 2-node workflow. Your task is to translate these codes to use the Parsl system. 
\end{shaded}\vspace*{40pt}}{\begin{myenv}{User prompt}
Task codes are provided below for the PyCOMPSs workflow system for a 2-node workflow. Your task is to translate these codes to use the Parsl system. 
\end{myenv}}

\subsection{Evaluation metrics}


Experiments we use in our study involve scientific code generation. We evaluate the LLMs by comparing against the reference codes as the ground truth using the following code similarity metrics:

\begin{itemize}
    \item \textbf{BLEU} metric~\cite{papineni2002bleu} was originally devised to evaluate machine-translated texts using n-gram precision measure.  We use the BLEU implementation provided by the sacrebleu package~\cite{post-2018-call}.

    \item \textbf{ChrF}~\cite{popovic2015chrf} is a character-based evaluation metric, where precision and recall are computed by averaging over character n-grams ranging from 1 to 6. We use the ChrF implementation provided by the sacrebleu package~\cite{post-2018-call}.
\end{itemize}

Both BLEU and ChrF scores range from 0.0 to 1.0, with higher values indicating better quality. Following common practice in the literature, we report these metrics multiplied by a factor of 100.




\section{Experiments}\label{sec:Experimental Methodology}


We evaluate the selected LLMs through a series of experiments including workflow configuration, task code annotation, and task code translation. We conduct our experiments on a workstation equipped with Apple M1 Max with 10-core CPU, 24-core GPU, and 32 GB unified memory. We use the Inspect AI framework\footnote{https://inspect.aisi.org.uk} for conducting our experiments, which are repeated over 5 times to mitigate the variability in the LLM responses.

In all of our experiments, we set the model $temperature$ to 0.2, and $top\_p$ to 0.95, which are commonly used in code generation experiments~\cite{nichols2024can} to produce more uniform distributions and to select a more representative sample of tokens from these distributions.\footnote{OpenAI's o-series reasoning models do not have $temperature$ and $top\_p$ parameters, hence, we apply these settings to all models except o3.}

\subsection{Workflow configuration}
\label{sec:wf-conf}
In this experiment, we use the LLMs to generate workflow configuration scripts and evaluate them using BLEU and ChrF metrics.  Note that we do not include PyCOMPSs and Parsl in this experiment because the configuration files in these two workflow systems describe the execution environment rather than the workflow structure.

\begin{table*}[h] 
\centering
\begin{small}
    \begin{tabular}{l|ll|ll|ll|ll|ll|}
\cline{2-11}
& \multicolumn{2}{c|}{\textbf{o3}}   & \multicolumn{2}{c|}{\textbf{Gemini-2.5-Pro}}     & \multicolumn{2}{c|}{\textbf{Claude-Sonnet-4}}   & \multicolumn{2}{c|}{\textbf{LLaMA-3.3-70B}}  & \multicolumn{2}{c|}{\textbf{Overall}} \\ \hline
\multicolumn{1}{|c|}{\textbf{Workflow systems}} 
& \multicolumn{1}{l|}{BLEU$\uparrow$} &  \multicolumn{1}{l|}{ChrF$\uparrow$} &  \multicolumn{1}{l|}{BLEU$\uparrow$} &  \multicolumn{1}{l|}{ChrF$\uparrow$}  &  \multicolumn{1}{l|}{BLEU$\uparrow$} &  \multicolumn{1}{l|}{ChrF$\uparrow$} &  \multicolumn{1}{l|}{BLEU$\uparrow$} &  \multicolumn{1}{l|}{ChrF$\uparrow$} &  \multicolumn{1}{l|}{BLEU$\uparrow$} &  \multicolumn{1}{l|}{ChrF$\uparrow$}
\\ \hline
\multicolumn{1}{|l|}{ADIOS2} & \multicolumn{1}{l|}{$59.1 \pm 2.3$} & \multicolumn{1}{l|}{$60.5 \pm 1.7$} & \multicolumn{1}{l|}{$73.0 \pm 1.8$} & \multicolumn{1}{l|}{$72.1 \pm 1.3$} & \multicolumn{1}{l|}{$72.1 \pm 0.0$} & \multicolumn{1}{l|}{$69.3 \pm 0.0$} & \multicolumn{1}{l|}{$35.9 \pm 0.7$} & \multicolumn{1}{l|}{$48.6 \pm 1.0$} & \multicolumn{1}{l|}{{\bf 60.0}\,${\boldmath\pm}$\,{\bf 3.5}} & \multicolumn{1}{l|}{{\bf 62.6}\,${\boldmath\pm}$\,{\bf 2.2}} \\ \hline
\multicolumn{1}{|l|}{Henson} & \multicolumn{1}{l|}{$20.2 \pm 2.3$} & \multicolumn{1}{l|}{$22.4 \pm 1.9$} & \multicolumn{1}{l|}{$26.9 \pm 1.9$} & \multicolumn{1}{l|}{$28.2 \pm 0.8$} & \multicolumn{1}{l|}{$25.0 \pm 0.0$} & \multicolumn{1}{l|}{$25.5 \pm 0.0$} & \multicolumn{1}{l|}{$27.7 \pm 1.0$} & \multicolumn{1}{l|}{$26.2 \pm 0.8$} & \multicolumn{1}{l|}{$25.0 \pm 1.0$} & \multicolumn{1}{l|}{$25.6 \pm 0.7$} \\ \hline
\multicolumn{1}{|l|}{Wilkins} & \multicolumn{1}{l|}{$30.0 \pm 1.5$} & \multicolumn{1}{l|}{$29.1 \pm 1.0$} & \multicolumn{1}{l|}{$31.6 \pm 3.4$} & \multicolumn{1}{l|}{$33.2 \pm 1.1$} & \multicolumn{1}{l|}{$36.8 \pm 0.8$} & \multicolumn{1}{l|}{$34.8 \pm 0.8$} & \multicolumn{1}{l|}{$39.0 \pm 0.0$} & \multicolumn{1}{l|}{$34.7 \pm 0.3$} & \multicolumn{1}{l|}{$34.4 \pm 1.2$} & \multicolumn{1}{l|}{$33.0 \pm 0.7$} \\ \hline
\multicolumn{1}{|l|}{Overall} & \multicolumn{1}{l|}{$36.5 \pm 4.5$} & \multicolumn{1}{l|}{$37.3 \pm 4.5$} & \multicolumn{1}{l|}{$43.8 \pm 5.7$} & \multicolumn{1}{l|}{{\bf 44.5}\,${\boldmath\pm}$\,{\bf 5.3}} & \multicolumn{1}{l|}{{\bf 44.6}\,${\boldmath\pm}$\,{\bf 5.3}} & \multicolumn{1}{l|}{$43.2 \pm 5.0$} & \multicolumn{1}{l|}{$34.2 \pm 1.3$} & \multicolumn{1}{l|}{$36.5 \pm 2.5$} & \multicolumn{1}{l|}{$39.8 \pm 2.3$} & \multicolumn{1}{l|}{$40.4 \pm 2.2$} \\ \hline
\end{tabular}
\end{small}
\caption{Evaluation of various LLMs using code similarity metrics for the task code configuration experiment. Average scores are displayed with a standard error from the mean over 5 runs.}
\label{tab:configuration}
\end{table*}

Table~\ref{tab:configuration} summarizes the results for each of the LLMs for different workflow systems. The mean score $\pm$ standard error over 5 trials are reported. Here, bold font in the overall row indicates the best-performing LLM model for this experiment, where we see that Gemini-2.5-Pro and Claude-Sonnet-4 outperform the other models. Bold font in the overall column highlights the workflow system where LLMs performed the best. We observe that this is the ADIOS2 system, while all of the evaluated LLMs \modif{demonstrate comparatively much lower performance on} Henson and Wilkins. We suspect this is mainly due to ADIOS2 being more widely used; hence, there are more documents (scientific papers, documentation) available online for training.

During this workflow configuration experiment, we find that LLMs often struggle to infer what configuration means for scientific workflows, and provide task code snippets instead of configuration files. We also observe frequent hallucinations. For instance, OpenAI model made up a document titled Wilkins Workflow System Documentation as a reference with the URL pointing to a personal blog website.\footnote{https://www.wilkins.io} 


\subsection{Task code annotation}


To evaluate the code annotation capabilities of LLMs, we provide a simple producer code written in C emulating an HPC simulation, adding comments where LLMs should annotate the code with relevant workflow system API calls. For PyCOMPSs and Parsl, we provide the equivalent of this code in Python, as these workflow systems are intended for Python task codes. We do not involve Wilkins in this study, as Wilkins does not require modifications to task codes.

Table~\ref{tab:annotation} shows the results for task code annotation capabilities of LLMs based on the code similarity metrics. In overall, we see that LLMs perform better compared with the workflow configuration experiment. This is because there are more code examples and documentation available online compared with the workflow configuration.

\begin{table*}[h] 
\centering
\begin{small}
    \begin{tabular}{l|ll|ll|ll|ll|ll|}
\cline{2-11}
& \multicolumn{2}{c|}{\textbf{o3}}   & \multicolumn{2}{c|}{\textbf{Gemini-2.5-Pro}}     & \multicolumn{2}{c|}{\textbf{Claude-Sonnet-4}}   & \multicolumn{2}{c|}{\textbf{LLaMA-3.3-70B}}  & \multicolumn{2}{c|}{\textbf{Overall}} \\ \hline
\multicolumn{1}{|c|}{\textbf{Workflow systems}} 
& \multicolumn{1}{l|}{BLEU$\uparrow$} &  \multicolumn{1}{l|}{ChrF$\uparrow$} &  \multicolumn{1}{l|}{BLEU$\uparrow$} &  \multicolumn{1}{l|}{ChrF$\uparrow$}  &  \multicolumn{1}{l|}{BLEU$\uparrow$} &  \multicolumn{1}{l|}{ChrF$\uparrow$} &  \multicolumn{1}{l|}{BLEU$\uparrow$} &  \multicolumn{1}{l|}{ChrF$\uparrow$} &  \multicolumn{1}{l|}{BLEU$\uparrow$} &  \multicolumn{1}{l|}{ChrF$\uparrow$}
\\ \hline
\multicolumn{1}{|l|}{ADIOS2} & \multicolumn{1}{l|}{$60.3 \pm 2.1$} & \multicolumn{1}{l|}{$59.0 \pm 1.7$} & \multicolumn{1}{l|}{$51.9 \pm 0.7$} & \multicolumn{1}{l|}{$54.7 \pm 1.5$} & \multicolumn{1}{l|}{$37.7 \pm 0.3$} & \multicolumn{1}{l|}{$34.1 \pm 0.1$} & \multicolumn{1}{l|}{$53.4 \pm 3.0$} & \multicolumn{1}{l|}{$55.9 \pm 2.0$} & \multicolumn{1}{l|}{$50.8 \pm 2.1$} & \multicolumn{1}{l|}{$50.9 \pm 2.4$} \\ \hline
\multicolumn{1}{|l|}{Henson} & \multicolumn{1}{l|}{$38.1 \pm 5.0$} & \multicolumn{1}{l|}{$36.1 \pm 4.2$} & \multicolumn{1}{l|}{$42.7 \pm 9.4$} & \multicolumn{1}{l|}{$47.1 \pm 8.7$} & \multicolumn{1}{l|}{$39.7 \pm 0.0$} & \multicolumn{1}{l|}{$49.7 \pm 0.9$} & \multicolumn{1}{l|}{$16.3 \pm 1.6$} & \multicolumn{1}{l|}{$19.6 \pm 1.5$} & \multicolumn{1}{l|}{$34.2 \pm 3.4$} & \multicolumn{1}{l|}{$38.1 \pm 3.5$} \\ \hline
\multicolumn{1}{|l|}{PyCOMPSs} & \multicolumn{1}{l|}{$72.4 \pm 1.8$} & \multicolumn{1}{l|}{$78.3 \pm 2.1$} & \multicolumn{1}{l|}{$89.3 \pm 3.1$} & \multicolumn{1}{l|}{$88.6 \pm 2.9$} & \multicolumn{1}{l|}{$49.7 \pm 0.0$} & \multicolumn{1}{l|}{$62.5 \pm 0.0$} & \multicolumn{1}{l|}{$9.9 \pm 4.0$} & \multicolumn{1}{l|}{$23.3 \pm 1.3$} & \multicolumn{1}{l|}{{\bf 55.5}\,${\boldmath\pm}$\,{\bf 7.0}} & \multicolumn{1}{l|}{{\bf 63.1}\,${\boldmath\pm}$\,{\bf 5.8}} \\ \hline
\multicolumn{1}{|l|}{Parsl} & \multicolumn{1}{l|}{$39.3 \pm 6.0$} & \multicolumn{1}{l|}{$57.1 \pm 2.4$} & \multicolumn{1}{l|}{$35.6 \pm 6.3$} & \multicolumn{1}{l|}{$55.2 \pm 4.2$} & \multicolumn{1}{l|}{$35.8 \pm 0.0$} & \multicolumn{1}{l|}{$49.7 \pm 0.0$} & \multicolumn{1}{l|}{$41.2 \pm 1.2$} & \multicolumn{1}{l|}{$57.2 \pm 0.1$} & \multicolumn{1}{l|}{$38.0 \pm 2.1$} & \multicolumn{1}{l|}{$54.8 \pm 1.3$} \\ \hline
\multicolumn{1}{|l|}{Overall} & \multicolumn{1}{l|}{$52.8 \pm 4.1$} & \multicolumn{1}{l|}{$57.7 \pm 3.7$} & \multicolumn{1}{l|}{{\bf 54.9}\,${\boldmath\pm}$\,{\bf 5.5}} & \multicolumn{1}{l|}{{\bf 61.4}\,${\boldmath\pm}$\,{\bf 4.4}} & \multicolumn{1}{l|}{$40.7 \pm 1.2$} & \multicolumn{1}{l|}{$49.0 \pm 2.3$} & \multicolumn{1}{l|}{$30.2 \pm 4.3$} & \multicolumn{1}{l|}{$39.0 \pm 4.1$} & \multicolumn{1}{l|}{$43.8 \pm 2.3$} & \multicolumn{1}{l|}{$50.8 \pm 2.1$} \\ \hline
\end{tabular}
\end{small}
\caption{Evaluation of various LLMs using code similarity metrics for the task code annotation experiment. Average scores are displayed with a standard error from the mean over 5 runs. }
\label{tab:annotation}
\end{table*}

As highlighted in the overall column, we observe that most of the LLMs provide better task annotations for the PyCOMPSs workflow system. This can be attributed to fact that annotations are at the core of PyCOMPSs' task-based workflow model, where users define the data and resource requirements of tasks by annotating Python methods. Interestingly, the LLaMa-3.3-70B model performs poorly for the PyCOMPSs system as its responses lack several required PyCOMPSs API calls such as $compss\_wait\_on\_file$, which is required for synchronization when communicating over files. 

Another interesting observation is that although the Parsl system is similar to PyCOMPSs in its use of task annotations, LLMs perform worse for Parsl---especially in terms of BLEU score. This performance drop is often due to redundant API calls in some models and missing required calls in others.  For instance, Parsl uses $Executor$ to control task-resource mapping and scheduling. In many of the generated code annotations, we see that LLMs configure Parsl with various type of executors even though this is not specified in the user prompt. As shown in Table~\ref{tab:annotation}, ChrF is more tolerant of such redundancies due to its character-based evaluation compared with BLEU, which is based on n-gram precision.



For the other workflow systems, we find that LLMs often use incorrect API calls. For example, Henson uses $henson\_save\_int$ for sending an integer variable. However, the o3 model generates $henson\_put$ while Gemini-2.5-Pro uses $henson\_declare\_variable$, which are both nonexistent.

\subsection{Task code translation}
In this experiment, we use LLMs to translate task codes between ADIOS2 and Henson workflow systems, and between PyCOMPSs and Parsl workflow systems. We provide the annotated producer codes from the previous experiment.



Table~\ref{tab:translation} shows the results for task code translation capabilities of LLMs based on the provided responses using the code similarity metrics. Overall, we observe that LLMs perform slightly worse than the task code annotation experiment for these individual workflow systems. We believe this is due to involving two different workflow systems, where LLMs need to have a knowledge of both of them. However, we obtain similar observations as in the annotation experiment. For example, LLMs perform better when translating the producer code to ADIOS2 compared with translating to Henson. Similarly, LLMs perform better when translating the producer code to PyCOMPSs compared with translating to Parsl.





Among the different LLMs, we see that the o3 model performs the best for the Henson to ADIOS2 translation, while Gemini-2.5-Pro outperforms the other evaluated models for the ADIOS2 to Henson translation. This demonstrates that no single model consistently performs best, even within the same experiment, as accuracy varies depending on the target workflow system.

To highlight the differences between the translated codes by different models, Table~\ref{tab:translate-adios} shows the codes generated by LLaMA-3.3-70B and Gemini-2.5-Pro for the ADIOS2 to Henson translation. Interestingly, LLaMA-3.3-70B produces Henson API calls as in the ADIOS2 API (e.g., $begin\_step$, $end\_step$, $declare\_var$, $put$), indicating that it lacks the knowledge of Henson system and mostly replaces $henson$ with $adios$ in the function names. On the other hand, Gemini-2.5-Pro  model uses correct Henson API calls for data exchange (lines 42 and 46), and issues the correct $henson\_yield$ (line 48) to signal the consumer that the data is ready, but it still hallucinates in several places and uses nonexistent API calls for creating Henson data types (lines 40-41 and 44-45), and for initialization (lines 10-12) and cleanup (line 54).






\begin{table*}[h] 
\centering
\begin{small}
    \begin{tabular}{l|ll|ll|ll|ll|ll|}
\cline{2-11}
& \multicolumn{2}{c|}{\textbf{o3}}   & \multicolumn{2}{c|}{\textbf{Gemini-2.5-Pro}}     & \multicolumn{2}{c|}{\textbf{Claude-Sonnet-4}}   & \multicolumn{2}{c|}{\textbf{LLaMA-3.3-70B}}  & \multicolumn{2}{c|}{\textbf{Overall}} \\ \hline
\multicolumn{1}{|c|}{\textbf{Workflow systems}} 
& \multicolumn{1}{l|}{BLEU$\uparrow$} &  \multicolumn{1}{l|}{ChrF$\uparrow$} &  \multicolumn{1}{l|}{BLEU$\uparrow$} &  \multicolumn{1}{l|}{ChrF$\uparrow$}  &  \multicolumn{1}{l|}{BLEU$\uparrow$} &  \multicolumn{1}{l|}{ChrF$\uparrow$} &  \multicolumn{1}{l|}{BLEU$\uparrow$} &  \multicolumn{1}{l|}{ChrF$\uparrow$} &  \multicolumn{1}{l|}{BLEU$\uparrow$} &  \multicolumn{1}{l|}{ChrF$\uparrow$}
\\ \hline
\multicolumn{1}{|l|}{Henson to ADIOS2} & \multicolumn{1}{l|}{$56.2 \pm 2.1$} & \multicolumn{1}{l|}{$54.8 \pm 1.4$} & \multicolumn{1}{l|}{$52.2 \pm 1.9$} & \multicolumn{1}{l|}{$49.3 \pm 1.7$} & \multicolumn{1}{l|}{$34.6 \pm 1.2$} & \multicolumn{1}{l|}{$33.1 \pm 1.2$} & \multicolumn{1}{l|}{$42.8 \pm 0.5$} & \multicolumn{1}{l|}{$45.9 \pm 0.7$} & \multicolumn{1}{l|}{$46.5 \pm 2.1$} & \multicolumn{1}{l|}{$45.8 \pm 1.9$} \\ \hline
\multicolumn{1}{|l|}{ADIOS2 to Henson} & \multicolumn{1}{l|}{$24.9 \pm 2.0$} & \multicolumn{1}{l|}{$39.6 \pm 1.8$} & \multicolumn{1}{l|}{$35.4 \pm 1.6$} & \multicolumn{1}{l|}{$50.2 \pm 1.6$} & \multicolumn{1}{l|}{$32.5 \pm 0.0$} & \multicolumn{1}{l|}{$40.6 \pm 0.1$} & \multicolumn{1}{l|}{$19.3 \pm 0.2$} & \multicolumn{1}{l|}{$30.2 \pm 0.3$} & \multicolumn{1}{l|}{$28.0 \pm 1.6$} & \multicolumn{1}{l|}{$40.1 \pm 1.7$} 
\\ \hline
\multicolumn{1}{|l|}{Parsl to PyCOMPSs} & \multicolumn{1}{l|}{$48.4 \pm 1.7$} & \multicolumn{1}{l|}{$70.6 \pm 2.1$} & \multicolumn{1}{l|}{$78.4 \pm 7.5$} & \multicolumn{1}{l|}{$82.3 \pm 5.4$} & \multicolumn{1}{l|}{$49.7 \pm 0.0$} & \multicolumn{1}{l|}{$62.5 \pm 0.0$} & \multicolumn{1}{l|}{$29.4 \pm 0.6$} & \multicolumn{1}{l|}{$42.1 \pm 1.5$} & \multicolumn{1}{l|}{{\bf 51.0}\,${\boldmath\pm}$\,{\bf 4.1}} & \multicolumn{1}{l|}{{\bf 61.3}\,${\boldmath\pm}$\,{\bf 4.0}} 
\\ \hline
\multicolumn{1}{|l|}{PyCOMPSs to Parsl} & \multicolumn{1}{l|}{$23.6 \pm 2.6$} & \multicolumn{1}{l|}{$48.5 \pm 2.5$} & \multicolumn{1}{l|}{$39.7 \pm 3.3$} & \multicolumn{1}{l|}{$60.2 \pm 1.7$} & \multicolumn{1}{l|}{$23.7 \pm 0.0$} & \multicolumn{1}{l|}{$57.1 \pm 0.0$} & \multicolumn{1}{l|}{$23.3 \pm 0.2$} & \multicolumn{1}{l|}{$44.4 \pm 0.1$} & \multicolumn{1}{l|}{$27.6 \pm 1.9$} & \multicolumn{1}{l|}{$52.5 \pm 1.6$} 
\\ \hline
\multicolumn{1}{|l|}{Overall} & \multicolumn{1}{l|}{$38.3 \pm 3.4$} & \multicolumn{1}{l|}{$53.4 \pm 4.9$} & \multicolumn{1}{l|}{{\bf 51.4}\,${\boldmath\pm}$\,{\bf 4.3}} & \multicolumn{1}{l|}{{\bf 60.5}\,${\boldmath\pm}$\,{\bf 3.4}} & \multicolumn{1}{l|}{$35.5 \pm 2.4$} & \multicolumn{1}{l|}{$47.6 \pm 3.3$} & \multicolumn{1}{l|}{$28.7 \pm 2.1$} & \multicolumn{1}{l|}{$40.7 \pm 1.5$} & \multicolumn{1}{l|}{$36.4 \pm 2.0$} & \multicolumn{1}{l|}{$48.2 \pm 1.9$} \\ \hline
\end{tabular}
\end{small}
\caption{Evaluation of various LLMs using code similarity metrics for the task code translation experiment. Average scores are displayed with a standard error from the mean over 5 runs.}
\label{tab:translation}
\end{table*}

\begin{table*}[htbp]
\centering
\begin{tabular}{|c|c|}
\hline
\begin{lstlisting}[
basicstyle={\tiny\ttfamily},
identifierstyle={\color{black}},
tabsize=2,
language={c},
numbersep=8pt,
numbers=left,
moredelim={[is][\textcolor{red}]{\^}{\^}}, 
xleftmargin=0.5cm,frame=tlbr,framesep=2pt,framerule=0pt
]  
#include <stdio.h>
#include <stdlib.h>
#include <unistd.h>
#include <time.h>
#include <mpi.h>
#include "henson.h"

int main(int argc, char** argv) {
    MPI_Init(&argc, &argv);
    int rank, size;
    MPI_Comm_rank(MPI_COMM_WORLD, &rank);
    MPI_Comm_size(MPI_COMM_WORLD, &size);

    size_t n = 50;
    if (argc > 1) n = atoi(argv[1]);
    if (rank == 0) printf("Using %zu random numbers\n", n);

    int iterations = 3;
    if (argc > 2) iterations = atoi(argv[2]);

    int sleep_interval = 0;
    if (argc > 3) sleep_interval = atoi(argv[3]);

    srand(time(NULL) + rank);

    ^henson_t h = henson_init(MPI_COMM_WORLD);
    henson_stage_t stage = henson_declare_stage(h, "SimulationOutput");

    henson_var_t varArray = henson_declare_var(stage, "array", HENSON_FLOAT, 2, 
    (size_t[]){size, n}, (size_t[]){rank, 0}, (size_t[]){1, n});
    henson_var_t varT = henson_declare_var(stage, "t", HENSON_INT, 0, 
    NULL, NULL, NULL);

    henson_output_t output = henson_open_output(stage, "output.bp", 
    HENSON_WRITE);^

    int t;
    for (t = 0; t < iterations; ++t) {
        if (sleep_interval) sleep(sleep_interval);

        float* array = malloc(n * sizeof(float));
        size_t i;
        for (i = 0; i < n; ++i) array[i] = (float) rand() / (float) RAND_MAX;

        float sum = 0;
        for (i = 0; i < n; ++i) sum += array[i];
        printf("[%d] Simulation [t=%d]: sum = %f\n", rank, t, sum); 

        float total_sum;
        MPI_Reduce(&sum, &total_sum, 1, MPI_FLOAT, MPI_SUM, 0, MPI_COMM_WORLD);
        if (rank == 0) 
        printf("[%d] Simulation [t=%d]: total_sum = %f\n", rank, t, total_sum); 

        ^henson_begin_step(output);
        henson_put_var(output, varArray, array);
        henson_put_var(output, varT, &t);
        henson_end_step(output);^

        free(array);
    }

    ^henson_close_output(output);^
    ^henson_finalize(h);^

    MPI_Finalize();
    return 0;
}
\end{lstlisting}
 
&
\begin{lstlisting}[
basicstyle={\tiny\ttfamily},
identifierstyle={\color{black}},
tabsize=2,
language={C},
numbersep=8pt,
numbers=left,
moredelim={[is][\textcolor{red}]{\^}{\^}}, 
xleftmargin=0.5cm,frame=tlbr,framesep=2pt,framerule=0pt
] 
#include <stdio.h>
#include <stdlib.h>
#include <unistd.h>
#include <time.h>
#include <mpi.h>
#include <henson/henson.h>

int main(int argc, char** argv)
{
    ^henson_init(argc, argv, MPI_COMM_WORLD);
    int rank = henson_rank();
    int size = henson_size();^

    size_t n = 50;
    if (argc > 1) n = atoi(argv[1]);
    if (rank == 0) printf("Using %zu random numbers\n", n);

    int sleep_interval = 0;
    if (argc > 2) sleep_interval = atoi(argv[2]);

    srand(time(NULL) + rank);

    int t = 0;
    while (henson_active())
    {
        if (sleep_interval) sleep(sleep_interval);

        float* array = (float*) malloc(n * sizeof(float));
        size_t i;
        for (i = 0; i < n; ++i) array[i] = (float) rand() / (float) RAND_MAX;

        float sum = 0;
        for (i = 0; i < n; ++i) sum += array[i];
        printf("[%d] Simulation [t=%d]: sum = %f\n", rank, t, sum); 

        float total_sum;
        MPI_Reduce(&sum, &total_sum, 1, MPI_FLOAT, MPI_SUM, 0, MPI_COMM_WORLD);
        if (rank == 0) 
            printf("[%d] Simulation [t=%d]: total_sum = %f\n", rank, t, total_sum); 

        ^henson_data_t array_hd;
        henson_data_init(&array_hd, HENSON_FLOAT, n, array);^
        henson_save("array", &array_hd);

        ^henson_data_t t_hd;
        henson_data_init_scalar(&t_hd, HENSON_INT, &t);^
        henson_save("t", &t_hd);

        henson_yield();

        free(array);
        t++;
    }

    ^henson_finalize();^
    return 0;
}
\end{lstlisting}    
\\
\hline
\end{tabular}
\caption{Translated producer codes with LLaMA-3.3-70B (left), and Gemini-2.5-Pro (right) for the Henson workflow system. Errors calling nonexistent API functions are indicated in red.} \label{tab:translate-adios} 
\end{table*}
\vspace{-0.1in}
\subsection{Prompt sensitivity}




In this experiment, we evaluate the sensitivity of LLMs to different prompting strategies. We develop five different prompts varying in style, wording, ordering, and amount of technical detail. We provide an example of these different prompting variations for task code annotation for the ADIOS2 workflow system below.


\aptLtoX{\begin{shaded}
\begin{center}\textbf{Prompt variations for task code annotation}\end{center}\par
\textbf{1. Original prompt}
You are assisting in the development of a simple producer-consumer workflow using the ADIOS2 system. The producer task code is provided below. Annotate this task code in order to use it with the ADIOS2 system.

\textbf{2. Different style}
Developer, please take the following producer task code and annotate it for compatibility with the ADIOS2 system in a producer-consumer workflow. Ensure all necessary ADIOS2 functions for data handling are included.

\textbf{3. Paraphrasing}
I have some code for a producer task that I want to integrate into a producer-consumer workflow using ADIOS2. Could you please go through the code provided below and add the necessary ADIOS2 annotations?

\textbf{4. Reordering}
Below is the producer task code for a simple producer-consumer workflow. Using the ADIOS2 system, please annotate this code to enable its use within the workflow.

\textbf{5. Detailed} 
Annotate the producer task code below with ADIOS2 calls (like DefineVariable, Put, BeginStep, EndStep) to enable it to run as part of an ADIOS2 workflow.
\end{shaded}\vskip10pt}{\begin{myenv}{Prompt variations for task code annotation}
\textbf{1. Original prompt}
You are assisting in the development of a simple producer-consumer workflow using the ADIOS2 system. The producer task code is provided below. Annotate this task code in order to use it with the ADIOS2 system.

\textbf{2. Different style}
Developer, please take the following producer task code and annotate it for compatibility with the ADIOS2 system in a producer-consumer workflow. Ensure all necessary ADIOS2 functions for data handling are included.

\textbf{3. Paraphrasing}
I have some code for a producer task that I want to integrate into a producer-consumer workflow using ADIOS2. Could you please go through the code provided below and add the necessary ADIOS2 annotations?

\textbf{4. Reordering}
Below is the producer task code for a simple producer-consumer workflow. Using the ADIOS2 system, please annotate this code to enable its use within the workflow.

\textbf{5. Detailed} 
Annotate the producer task code below with ADIOS2 calls (like DefineVariable, Put, BeginStep, EndStep) to enable it to run as part of an ADIOS2 workflow.
\end{myenv}}

\begin{figure*}[h]
        \includegraphics{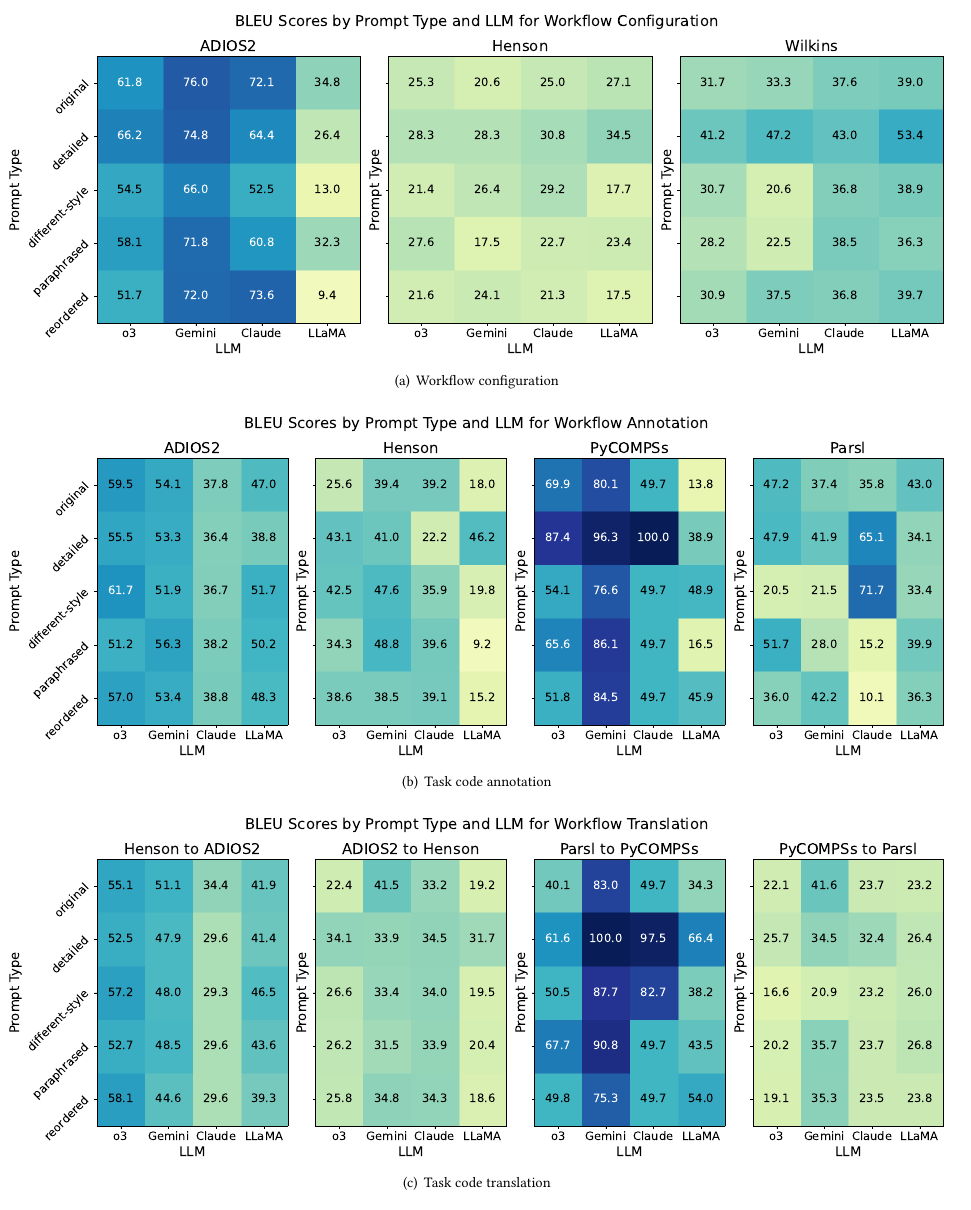}
        \caption{Prompt sensitivity results for workflow experiments. We see that there is no superior prompt across all LLMs.}
        \label{fig:prompt-sensitivity}
\end{figure*}




\vspace{0.2in}
Figure~\ref{fig:prompt-sensitivity} shows the obtained BLEU scores corresponding to different prompting strategies for all the workflow experiments including configuration, task code annotation, and translation. ChrF scores are similar to BLEU scores, and are omitted to simplify the presentation of results. The results show that there is no single prompt strategy consistently performing best across all LLMs. Instead, we see that different prompts yield better results depending on the model that is being used. We believe that this variability is due to the differences in model training and their instruction styles, which makes it difficult to construct a single effective prompt. For instance, one might expect detailed prompts to perform better as they include technical details for workflow systems' API. However, we see that this is not always the case, indicating that LLMs are not able to benefit from these details without the proper examples showing the usage of these API calls.

Despite the variations, many of our observations still remain. For example, for the task code annotation experiment, LLMs perform better when annotating the task codes for the PyCOMPSs workflow system, as shown in Figure 1(b). Moreover, in Figure 1(a) we see the least variability among prompts in the workflow configuration experiment for Henson and Wilkins systems, as all LLMs struggle with configuring these workflow systems, resulting in low scores for most of the prompt variations.

\subsection{Few-shot prompting}


\begin{table*}[h] 
\centering
\begin{small}

    \begin{tabular}{l|ll|ll|ll|ll|l|l|}
\cline{2-9}
& \multicolumn{2}{c|}{\textbf{o3}}   & \multicolumn{2}{c|}{\textbf{Gemini-2.5-Pro}}     & \multicolumn{2}{c|}{\textbf{Claude-Sonnet-4}}   & \multicolumn{2}{c|}{\textbf{LLaMA-3.3-70B}}  \\ \hline
\multicolumn{1}{|c|}{\textbf{Approach}} 
& \multicolumn{1}{l|}{BLEU$\uparrow$} &  \multicolumn{1}{l|}{ChrF$\uparrow$} &  \multicolumn{1}{l|}{BLEU$\uparrow$} &  \multicolumn{1}{l|}{ChrF$\uparrow$}  &  \multicolumn{1}{l|}{BLEU$\uparrow$} &  \multicolumn{1}{l|}{ChrF$\uparrow$} &  \multicolumn{1}{l|}{BLEU$\uparrow$} &  \multicolumn{1}{l|}{ChrF$\uparrow$}
\\ \hline
\multicolumn{1}{|l|}{Original (zero-shot)}                                                              & \multicolumn{1}{l|}{$36.5 \pm 4.5$}    & \multicolumn{1}{l|}{$37.3 \pm 4.5$}          & \multicolumn{1}{l|}{$43.8 \pm 5.7$}           & \multicolumn{1}{l|}{$44.5 \pm 5.3$} &
\multicolumn{1}{l|}{$44.6 \pm 5.3$}    & \multicolumn{1}{l|}{$43.2 \pm 5.0$}          & \multicolumn{1}{l|}{$34.2 \pm 1.3$}           & \multicolumn{1}{l|}{$36.5 \pm 2.5$}
\\ \hline
\multicolumn{1}{|l|}{Few-shot prompting}                                                              & 
\multicolumn{1}{l|}{{\bf 89.3}\,${\boldmath\pm}$\,{\bf 2.7}}    &  \multicolumn{1}{l|}{{\bf 89.7}\,${\boldmath\pm}$\,{\bf 2.6}}         & \multicolumn{1}{l|}{{\bf 86.7}\,${\boldmath\pm}$\,{\bf 2.3}}         & \multicolumn{1}{l|}{{\bf 87.6}\,${\boldmath\pm}$\,{\bf 2.1}}      &
\multicolumn{1}{l|}{{\bf 91.5}\,${\boldmath\pm}$\,{\bf 3.0}}    &  \multicolumn{1}{l|}{{\bf 95.9}\,${\boldmath\pm}$\,{\bf 2.4}}         & \multicolumn{1}{l|}{{\bf 84.1}\,${\boldmath\pm}$\,{\bf 2.1}}         & \multicolumn{1}{l|}{{\bf 85.0}\,${\boldmath\pm}$\,{\bf 2.4}}  

\\ \hline

\end{tabular}
\end{small}
\caption{Comparison of few-shot prompting vs. zero-shot prompting for the workflow configuration experiment. The BLEU and ChrF scores are averaged over the different workflow systems used.}
\label{tab:few-shot}
\end{table*}





As a first step toward extending LLM capabilities for scientific workflows, we explore the few-shot prompting technique for the workflow configuration experiment. We chose this experiment since this is where LLMs struggled the most. For few-shot prompting, we augment the original user prompt in Section~\ref{sec:wf-conf} with an example configuration file for a simple 2-node workflow. Table~\ref{tab:few-shot} shows the obtained code similarity metrics corresponding to original and few-shot prompting approaches averaged over the different workflow systems for the workflow configuration experiment. In Table~\ref{tab:few-shot}, the bold font denotes the top score in each column, indicating the better approach for each LLM model. We can see that few-shot prompting increases the quality of the generated configuration scripts for all of the evaluated LLMs, with Claude-Sonnet-4 attaining the most gain. We find that providing examples helps the LLMs avoid generating configuration code that does not exist in the workflow systems, thereby preventing hallucinations and increasing the response quality.

In order to highlight the importance of providing workflow-specific context via few-shot prompting, we report the generated configuration scripts with and without few-shot prompting for Wilkins using o3 in Table~\ref{tab:config-wilkins}. With few-shot prompting, o3 generates the correct configuration file. Without the provided context, o3 uses nonexisting fields like inputs and outputs instead of inports and outports to define the data requirements. This is expected as inports/outports are specific to Wilkins, while input/output are commonly used in the literature. Similarly, o3 uses command and dependencies fields which are nonexistent in the Wilkins system.

\begin{table*}[htbp]
\centering
\begin{tabular}{|p{20pc}|p{20pc}|}
\hline
\begin{lstlisting}[
basicstyle={\tiny\ttfamily},
identifierstyle={\color{black}},
tabsize=2,
language={C},
numbersep=8pt,
numbers=left,
xleftmargin=0.5cm,frame=tlbr,framesep=2pt,framerule=0pt
]         
tasks:
  - func: producer
    nprocs: 3
    outports:
      - filename: outfile.h5
        dsets:
          - name: /group1/grid
            file: 0
            memory: 1
          - name: /group1/particles
            file: 0
            memory: 1
  - func: consumer1
    nprocs: 1
    inports:
      - filename: outfile.h5
        dsets:
          - name: /group1/grid
            file: 0
            memory: 1
  - func: consumer2
    nprocs: 1
    inports:
      - filename: outfile.h5
        dsets:
          - name: /group1/particles
            file: 0
            memory: 1
\end{lstlisting}

&
\begin{lstlisting}[
basicstyle={\tiny\ttfamily},
identifierstyle={\color{black}},
tabsize=2,
language={C},
numbersep=8pt,
numbers=left,
moredelim={[is][\textcolor{red}]{\^}{\^}}, 
xleftmargin=0.5cm,frame=tlbr,framesep=2pt,framerule=0pt
] 
# wilkins_workflow.yaml

^workflow:
  name: simple_3node_workflow
  datasets:
    grid: {}
    particles: {}^
  tasks:
    producer:
      ^command: ./producer^
      ^processes:^ 3
      ^outputs:^
        - grid
        - particles
    consumer1:
      ^command: ./consumer_grid^
      ^processes:^ 1
      ^inputs:^
        - grid
    consumer2:
      ^command: ./consumer_particles^
     ^ processes:^ 1
      ^inputs:^
        - particles
^dependencies:
  - from: producer
    to: consumer1
    datasets:
      - grid
  - from: producer
    to: consumer2
    datasets:
      - particles^
\end{lstlisting}    
\\
\hline
\end{tabular}
\caption{Generated configuration files with few-shot prompting (left), and with zero-shot prompting (right) using o3 for the Wilkins workflow system.} \label{tab:config-wilkins}
\end{table*}




\subsection{Summary and discussion}


In our experiments, we observe that LLMs lack understanding of scientific workflows, often leading to incorrect assumptions and hallucinations. We believe that addressing this limitation is key to improving the applicability of LLMs to scientific workflows. This can be achieved by providing the contextual knowledge of scientific workflows and workflow systems to LLMs by applying several strategies, including few-shot prompting, retrieval-augmented generation (RAG), and fine-tuning. We demonstrate the effectiveness of providing such context via few-shot prompting for the workflow configuration experiment with the o3 model. 


Overall, we observe that the performance of LLMs varies across workflow systems. The evaluated LLMs often perform better with workflows involving well-documented systems such as ADIOS2 and PyCOMPSs, likely due to the availability of relevant documentation in their training data. 

Lastly, we observe that the performance of LLMs varies across experiments. For example, Claude-Sonnet-4 performs well for configuring the ADIOS2 system, but \modif{Claude-Sonnet-4 shows weaker performance} in the workflow annotation experiment for ADIOS2. In contrast, LLaMA-3.3-70B shows better performance in annotating task code with ADIOS2 API calls than in configuring the ADIOS2 system. This suggests that different models may be better suited to specific aspects of scientific workflows.

\section{Conclusion}\label{sec:Conclusion}

Scientific workflows play an important role in the scientific pipeline from hypothesis to discovery. While they offer advanced capabilities, challenges related to their usability and explainability remain. Recent advancements in LLMs present opportunities to address these issues. In this paper, we conduct an extensive experimental campaign evaluating capabilities of various LLMs across different workflow experiments and systems. Our results reveal that LLMs \modif{face challenges} in the context of scientific workflows, as LLMs are \modif{primarily} trained for general tasks and lack the domain knowledge required for scientific workflows. However, our results also show the promising potential of LLMs, highlighting the need for further research to address their limitations. For instance, we observe that incorporating external knowledge as context significantly improves the performance of LLMs, as we demonstrated with the evaluated models in the few-shot prompting experiment for the workflow configuration. Another promising direction for advancing LLM capabilities in scientific workflows is to incorporate iterative error correction mechanisms as successfully applied in other LLM applications to scientific code generation~\cite{peterka2025chatvis, mallick2024chatvis}.

\begin{acks}
This material is based upon work supported by the U.S. Department of Energy, Office 
of Science, Office of Advanced Scientific Computing Research, under contract numbers DE-AC02-06CH11357, program manager Hal Finkel.
\end{acks}

\bibliographystyle{ACM-Reference-Format}
\bibliography{biblio}


\begin{thebibliography}{33}


\ifx \showCODEN    \undefined \def \showCODEN     #1{\unskip}     \fi
\ifx \showDOI      \undefined \def \showDOI       #1{#1}\fi
\ifx \showISBNx    \undefined \def \showISBNx     #1{\unskip}     \fi
\ifx \showISBNxiii \undefined \def \showISBNxiii  #1{\unskip}     \fi
\ifx \showISSN     \undefined \def \showISSN      #1{\unskip}     \fi
\ifx \showLCCN     \undefined \def \showLCCN      #1{\unskip}     \fi
\ifx \shownote     \undefined \def \shownote      #1{#1}          \fi
\ifx \showarticletitle \undefined \def \showarticletitle #1{#1}   \fi
\ifx \showURL      \undefined \def \showURL       {\relax}        \fi
\providecommand\bibfield[2]{#2}
\providecommand\bibinfo[2]{#2}
\providecommand\natexlab[1]{#1}
\providecommand\showeprint[2][]{arXiv:#2}

\bibitem[Achiam et~al\mbox{.}(2023)]%
        {achiam2023gpt}
\bibfield{author}{\bibinfo{person}{Josh Achiam}, \bibinfo{person}{Steven
  Adler}, \bibinfo{person}{Sandhini Agarwal}, \bibinfo{person}{Lama Ahmad},
  \bibinfo{person}{Ilge Akkaya}, \bibinfo{person}{Florencia~Leoni Aleman},
  \bibinfo{person}{Diogo Almeida}, \bibinfo{person}{Janko Altenschmidt},
  \bibinfo{person}{Sam Altman}, \bibinfo{person}{Shyamal Anadkat},
  {et~al\mbox{.}}} \bibinfo{year}{2023}\natexlab{}.
\newblock \showarticletitle{Gpt-4 technical report}.
\newblock \bibinfo{journal}{\emph{arXiv preprint arXiv:2303.08774}}
  (\bibinfo{year}{2023}).
\newblock


\bibitem[Altintas et~al\mbox{.}(2019)]%
        {altintas2019toward}
\bibfield{author}{\bibinfo{person}{Ilkay Altintas}, \bibinfo{person}{Shweta
  Purawat}, \bibinfo{person}{Daniel Crawl}, \bibinfo{person}{Alok Singh}, {and}
  \bibinfo{person}{Kyle Marcus}.} \bibinfo{year}{2019}\natexlab{}.
\newblock \showarticletitle{Toward a methodology and framework for
  workflow-driven team science}.
\newblock \bibinfo{journal}{\emph{Computing in Science \& Engineering}}
  \bibinfo{volume}{21}, \bibinfo{number}{4} (\bibinfo{year}{2019}),
  \bibinfo{pages}{37--48}.
\newblock


\bibitem[Anil et~al\mbox{.}(2023)]%
        {anil2023palm}
\bibfield{author}{\bibinfo{person}{Rohan Anil}, \bibinfo{person}{Andrew~M Dai},
  \bibinfo{person}{Orhan Firat}, \bibinfo{person}{Melvin Johnson},
  \bibinfo{person}{Dmitry Lepikhin}, \bibinfo{person}{Alexandre Passos},
  \bibinfo{person}{Siamak Shakeri}, \bibinfo{person}{Emanuel Taropa},
  \bibinfo{person}{Paige Bailey}, \bibinfo{person}{Zhifeng Chen},
  {et~al\mbox{.}}} \bibinfo{year}{2023}\natexlab{}.
\newblock \showarticletitle{Palm 2 technical report}.
\newblock \bibinfo{journal}{\emph{arXiv preprint arXiv:2305.10403}}
  (\bibinfo{year}{2023}).
\newblock


\bibitem[Ayachit et~al\mbox{.}(2015)]%
        {ayachit2015paraview}
\bibfield{author}{\bibinfo{person}{Utkarsh Ayachit}, \bibinfo{person}{Andrew
  Bauer}, \bibinfo{person}{Berk Geveci}, \bibinfo{person}{Patrick O'Leary},
  \bibinfo{person}{Kenneth Moreland}, \bibinfo{person}{Nathan Fabian}, {and}
  \bibinfo{person}{Jeffrey Mauldin}.} \bibinfo{year}{2015}\natexlab{}.
\newblock \showarticletitle{{ParaView} {C}atalyst: Enabling in situ data
  analysis and visualization}. In \bibinfo{booktitle}{\emph{Proceedings of the
  First Workshop on In Situ Infrastructures for Enabling Extreme-Scale Analysis
  and Visualization}}. ACM, \bibinfo{pages}{25--29}.
\newblock


\bibitem[Ayachit et~al\mbox{.}(2016)]%
        {ayachit2016sensei}
\bibfield{author}{\bibinfo{person}{Utkarsh Ayachit}, \bibinfo{person}{Brad
  Whitlock}, \bibinfo{person}{Matthew Wolf}, \bibinfo{person}{Burlen Loring},
  \bibinfo{person}{Berk Geveci}, \bibinfo{person}{David Lonie}, {and}
  \bibinfo{person}{E Bethel}.} \bibinfo{year}{2016}\natexlab{}.
\newblock \showarticletitle{The {SENSEI} generic in situ interface}. In
  \bibinfo{booktitle}{\emph{Proceedings of the 2nd Workshop on In Situ
  Infrastructures for Enabling Extreme-scale Analysis and Visualization}}. IEEE
  Press, \bibinfo{pages}{40--44}.
\newblock


\bibitem[Babuji et~al\mbox{.}(2019)]%
        {babuji2019parsl}
\bibfield{author}{\bibinfo{person}{Yadu Babuji}, \bibinfo{person}{Anna
  Woodard}, \bibinfo{person}{Zhuozhao Li}, \bibinfo{person}{Daniel~S Katz},
  \bibinfo{person}{Ben Clifford}, \bibinfo{person}{Rohan Kumar},
  \bibinfo{person}{Lukasz Lacinski}, \bibinfo{person}{Ryan Chard},
  \bibinfo{person}{Justin~M Wozniak}, \bibinfo{person}{Ian Foster},
  {et~al\mbox{.}}} \bibinfo{year}{2019}\natexlab{}.
\newblock \showarticletitle{Parsl: Pervasive parallel programming in python}.
  In \bibinfo{booktitle}{\emph{Proceedings of the 28th International Symposium
  on High-Performance Parallel and Distributed Computing}}.
  \bibinfo{pages}{25--36}.
\newblock


\bibitem[Boyuka et~al\mbox{.}(2014)]%
        {boyuka2014transparent}
\bibfield{author}{\bibinfo{person}{David~A Boyuka}, \bibinfo{person}{Sriram
  Lakshminarasimham}, \bibinfo{person}{Xiaocheng Zou},
  \bibinfo{person}{Zhenhuan Gong}, \bibinfo{person}{John Jenkins},
  \bibinfo{person}{Eric~R Schendel}, \bibinfo{person}{Norbert Podhorszki},
  \bibinfo{person}{Qing Liu}, \bibinfo{person}{Scott Klasky}, {and}
  \bibinfo{person}{Nagiza~F Samatova}.} \bibinfo{year}{2014}\natexlab{}.
\newblock \showarticletitle{Transparent in situ data transformations in adios}.
  In \bibinfo{booktitle}{\emph{2014 14th IEEE/ACM International Symposium on
  Cluster, Cloud and Grid Computing}}. IEEE, \bibinfo{pages}{256--266}.
\newblock


\bibitem[Chen et~al\mbox{.}(2023)]%
        {chen2023lm4hpc}
\bibfield{author}{\bibinfo{person}{Le Chen}, \bibinfo{person}{Pei-Hung Lin},
  \bibinfo{person}{Tristan Vanderbruggen}, \bibinfo{person}{Chunhua Liao},
  \bibinfo{person}{Murali Emani}, {and} \bibinfo{person}{Bronis de Supinski}.}
  \bibinfo{year}{2023}\natexlab{}.
\newblock \showarticletitle{Lm4hpc: Towards effective language model
  application in high-performance computing}. In
  \bibinfo{booktitle}{\emph{International Workshop on OpenMP}}. Springer,
  \bibinfo{pages}{18--33}.
\newblock


\bibitem[Deelman et~al\mbox{.}(2009)]%
        {deelman2009workflows}
\bibfield{author}{\bibinfo{person}{Ewa Deelman}, \bibinfo{person}{Dennis
  Gannon}, \bibinfo{person}{Matthew Shields}, {and} \bibinfo{person}{Ian
  Taylor}.} \bibinfo{year}{2009}\natexlab{}.
\newblock \showarticletitle{Workflows and e-Science: An overview of workflow
  system features and capabilities}.
\newblock \bibinfo{journal}{\emph{Future generation computer systems}}
  \bibinfo{volume}{25}, \bibinfo{number}{5} (\bibinfo{year}{2009}),
  \bibinfo{pages}{528--540}.
\newblock


\bibitem[Deelman et~al\mbox{.}(2019)]%
        {deelman2019evolution}
\bibfield{author}{\bibinfo{person}{Ewa Deelman}, \bibinfo{person}{Karan Vahi},
  \bibinfo{person}{Mats Rynge}, \bibinfo{person}{Rajiv Mayani},
  \bibinfo{person}{Rafael~Ferreira da Silva}, \bibinfo{person}{George
  Papadimitriou}, {and} \bibinfo{person}{Miron Livny}.}
  \bibinfo{year}{2019}\natexlab{}.
\newblock \showarticletitle{The evolution of the pegasus workflow management
  software}.
\newblock \bibinfo{journal}{\emph{Computing in Science \& Engineering}}
  \bibinfo{volume}{21}, \bibinfo{number}{4} (\bibinfo{year}{2019}),
  \bibinfo{pages}{22--36}.
\newblock


\bibitem[Dorier et~al\mbox{.}(2016)]%
        {dorier2016damaris}
\bibfield{author}{\bibinfo{person}{Matthieu Dorier}, \bibinfo{person}{Gabriel
  Antoniu}, \bibinfo{person}{Franck Cappello}, \bibinfo{person}{Marc Snir},
  \bibinfo{person}{Robert Sisneros}, \bibinfo{person}{Orcun Yildiz},
  \bibinfo{person}{Shadi Ibrahim}, \bibinfo{person}{Tom Peterka}, {and}
  \bibinfo{person}{Leigh Orf}.} \bibinfo{year}{2016}\natexlab{}.
\newblock \showarticletitle{Damaris: Addressing performance variability in data
  management for post-petascale simulations}.
\newblock \bibinfo{journal}{\emph{ACM Transactions on Parallel Computing
  (TOPC)}} \bibinfo{volume}{3}, \bibinfo{number}{3} (\bibinfo{year}{2016}),
  \bibinfo{pages}{15}.
\newblock


\bibitem[Duque et~al\mbox{.}(2023)]%
        {duque2023leveraging}
\bibfield{author}{\bibinfo{person}{Alejandro Duque}, \bibinfo{person}{Abdullah
  Syed}, \bibinfo{person}{Kastan~V Day}, \bibinfo{person}{Matthew~J Berry},
  \bibinfo{person}{Daniel~S Katz}, {and} \bibinfo{person}{Volodymyr~V
  Kindratenko}.} \bibinfo{year}{2023}\natexlab{}.
\newblock \showarticletitle{Leveraging Large Language Models to Build and
  Execute Computational Workflows}.
\newblock \bibinfo{journal}{\emph{arXiv preprint arXiv:2312.07711}}
  (\bibinfo{year}{2023}).
\newblock


\bibitem[Gueroudji et~al\mbox{.}(2021)]%
        {gueroudji2021deisa}
\bibfield{author}{\bibinfo{person}{Amal Gueroudji}, \bibinfo{person}{Julien
  Bigot}, {and} \bibinfo{person}{Bruno Raffin}.}
  \bibinfo{year}{2021}\natexlab{}.
\newblock \showarticletitle{DEISA: dask-enabled in situ analytics}. In
  \bibinfo{booktitle}{\emph{2021 IEEE 28th International Conference on High
  Performance Computing, Data, and Analytics (HiPC)}}. IEEE,
  \bibinfo{pages}{11--20}.
\newblock


\bibitem[Kuhlen et~al\mbox{.}(2011)]%
        {kuhlen2011parallel}
\bibfield{author}{\bibinfo{person}{T Kuhlen}, \bibinfo{person}{R Pajarola},
  {and} \bibinfo{person}{K Zhou}.} \bibinfo{year}{2011}\natexlab{}.
\newblock \showarticletitle{Parallel in situ coupling of simulation with a
  fully featured visualization system}. In
  \bibinfo{booktitle}{\emph{Proceedings of the 11th Eurographics Conference on
  Parallel Graphics and Visualization (EGPGV)}}.
\newblock


\bibitem[Kumar et~al\mbox{.}(2023)]%
        {kumar2023mycrunchgpt}
\bibfield{author}{\bibinfo{person}{Varun Kumar}, \bibinfo{person}{Leonard
  Gleyzer}, \bibinfo{person}{Adar Kahana}, \bibinfo{person}{Khemraj Shukla},
  {and} \bibinfo{person}{George~Em Karniadakis}.}
  \bibinfo{year}{2023}\natexlab{}.
\newblock \showarticletitle{Mycrunchgpt: A llm assisted framework for
  scientific machine learning}.
\newblock \bibinfo{journal}{\emph{Journal of Machine Learning for Modeling and
  Computing}} \bibinfo{volume}{4}, \bibinfo{number}{4} (\bibinfo{year}{2023}).
\newblock


\bibitem[Mallick et~al\mbox{.}(2024)]%
        {mallick2024chatvis}
\bibfield{author}{\bibinfo{person}{Tanwi Mallick}, \bibinfo{person}{Orcun
  Yildiz}, \bibinfo{person}{David Lenz}, {and} \bibinfo{person}{Tom Peterka}.}
  \bibinfo{year}{2024}\natexlab{}.
\newblock \showarticletitle{ChatVis: Automating Scientific Visualization with a
  Large Language Model}.
\newblock \bibinfo{journal}{\emph{arXiv preprint arXiv:2410.11863}}
  (\bibinfo{year}{2024}).
\newblock


\bibitem[Morozov and Lukic(2016)]%
        {morozov2016master}
\bibfield{author}{\bibinfo{person}{Dmitriy Morozov} {and}
  \bibinfo{person}{Zarija Lukic}.} \bibinfo{year}{2016}\natexlab{}.
\newblock \showarticletitle{Master of puppets: {C}ooperative multitasking for
  in situ processing}. In \bibinfo{booktitle}{\emph{Proceedings of the 25th ACM
  International Symposium on High-Performance Parallel and Distributed
  Computing}}. ACM, \bibinfo{pages}{285--288}.
\newblock


\bibitem[Nichols et~al\mbox{.}(2024a)]%
        {nichols2024can}
\bibfield{author}{\bibinfo{person}{Daniel Nichols}, \bibinfo{person}{Joshua~H
  Davis}, \bibinfo{person}{Zhaojun Xie}, \bibinfo{person}{Arjun Rajaram}, {and}
  \bibinfo{person}{Abhinav Bhatele}.} \bibinfo{year}{2024}\natexlab{a}.
\newblock \showarticletitle{Can large language models write parallel code?}. In
  \bibinfo{booktitle}{\emph{Proceedings of the 33rd International Symposium on
  High-Performance Parallel and Distributed Computing}}.
  \bibinfo{pages}{281--294}.
\newblock


\bibitem[Nichols et~al\mbox{.}(2024b)]%
        {nichols2024hpc}
\bibfield{author}{\bibinfo{person}{Daniel Nichols}, \bibinfo{person}{Aniruddha
  Marathe}, \bibinfo{person}{Harshitha Menon}, \bibinfo{person}{Todd Gamblin},
  {and} \bibinfo{person}{Abhinav Bhatele}.} \bibinfo{year}{2024}\natexlab{b}.
\newblock \showarticletitle{HPC-Coder: Modeling Parallel Programs using Large
  Language Models}. In \bibinfo{booktitle}{\emph{ISC High Performance 2024
  Research Paper Proceedings (39th International Conference)}}. Prometeus GmbH,
  \bibinfo{pages}{1--12}.
\newblock


\bibitem[OpenAI(2025)]%
        {jaech2024openai}
\bibfield{author}{\bibinfo{person}{OpenAI}.} \bibinfo{year}{2025}\natexlab{}.
\newblock \bibinfo{title}{Openai o3 and o4-mini system card}.
\newblock
  \bibinfo{howpublished}{\url{https://openai.com/index/o3-o4-mini-system-card/}}.
\newblock
\newblock
\shownote{Accessed: 2025-07-30}.


\bibitem[Papineni et~al\mbox{.}(2002)]%
        {papineni2002bleu}
\bibfield{author}{\bibinfo{person}{Kishore Papineni}, \bibinfo{person}{Salim
  Roukos}, \bibinfo{person}{Todd Ward}, {and} \bibinfo{person}{Wei-Jing Zhu}.}
  \bibinfo{year}{2002}\natexlab{}.
\newblock \showarticletitle{Bleu: a method for automatic evaluation of machine
  translation}. In \bibinfo{booktitle}{\emph{Proceedings of the 40th annual
  meeting of the Association for Computational Linguistics}}.
  \bibinfo{pages}{311--318}.
\newblock


\bibitem[Peterka et~al\mbox{.}(2025)]%
        {peterka2025chatvis}
\bibfield{author}{\bibinfo{person}{Tom Peterka}, \bibinfo{person}{Tanwi
  Mallick}, \bibinfo{person}{Orcun Yildiz}, \bibinfo{person}{David Lenz},
  \bibinfo{person}{Cory Quammen}, {and} \bibinfo{person}{Berk Geveci}.}
  \bibinfo{year}{2025}\natexlab{}.
\newblock \showarticletitle{ChatVis: Large Language Model Agent for Generating
  Scientific Visualizations}.
\newblock \bibinfo{journal}{\emph{arXiv preprint arXiv:2507.23096}}
  (\bibinfo{year}{2025}).
\newblock


\bibitem[Popovi{\'c}(2015)]%
        {popovic2015chrf}
\bibfield{author}{\bibinfo{person}{Maja Popovi{\'c}}.}
  \bibinfo{year}{2015}\natexlab{}.
\newblock \showarticletitle{chrF: character n-gram F-score for automatic MT
  evaluation}. In \bibinfo{booktitle}{\emph{Proceedings of the tenth workshop
  on statistical machine translation}}. \bibinfo{pages}{392--395}.
\newblock


\bibitem[Post(2018)]%
        {post-2018-call}
\bibfield{author}{\bibinfo{person}{Matt Post}.}
  \bibinfo{year}{2018}\natexlab{}.
\newblock \showarticletitle{A Call for Clarity in Reporting {BLEU} Scores}. In
  \bibinfo{booktitle}{\emph{Proceedings of the Third Conference on Machine
  Translation: Research Papers}}. \bibinfo{publisher}{Association for
  Computational Linguistics}, \bibinfo{address}{Belgium, Brussels},
  \bibinfo{pages}{186--191}.
\newblock
\urldef\tempurl%
\url{https://www.aclweb.org/anthology/W18-6319}
\showURL{%
\tempurl}


\bibitem[S{\"a}nger et~al\mbox{.}(2023)]%
        {sanger2023large}
\bibfield{author}{\bibinfo{person}{Mario S{\"a}nger}, \bibinfo{person}{Ninon
  De~Mecquenem}, \bibinfo{person}{Katarzyna~Ewa Lewi{\'n}ska},
  \bibinfo{person}{Vasilis Bountris}, \bibinfo{person}{Fabian Lehmann},
  \bibinfo{person}{Ulf Leser}, {and} \bibinfo{person}{Thomas Kosch}.}
  \bibinfo{year}{2023}\natexlab{}.
\newblock \showarticletitle{Large Language Models to the Rescue: Reducing the
  Complexity in Scientific Workflow Development Using ChatGPT}.
\newblock \bibinfo{journal}{\emph{arXiv preprint arXiv:2311.01825}}
  (\bibinfo{year}{2023}).
\newblock


\bibitem[Tejedor et~al\mbox{.}(2017)]%
        {tejedor2017pycompss}
\bibfield{author}{\bibinfo{person}{Enric Tejedor}, \bibinfo{person}{Yolanda
  Becerra}, \bibinfo{person}{Guillem Alomar}, \bibinfo{person}{Anna Queralt},
  \bibinfo{person}{Rosa~M Badia}, \bibinfo{person}{Jordi Torres},
  \bibinfo{person}{Toni Cortes}, {and} \bibinfo{person}{Jes{\'u}s Labarta}.}
  \bibinfo{year}{2017}\natexlab{}.
\newblock \showarticletitle{{PyCOMPSs}: Parallel computational workflows in
  {P}ython}.
\newblock \bibinfo{journal}{\emph{The International Journal of High Performance
  Computing Applications}} \bibinfo{volume}{31}, \bibinfo{number}{1}
  (\bibinfo{year}{2017}), \bibinfo{pages}{66--82}.
\newblock


\bibitem[Thoppilan et~al\mbox{.}(2022)]%
        {thoppilan2022lamda}
\bibfield{author}{\bibinfo{person}{Romal Thoppilan}, \bibinfo{person}{Daniel
  De~Freitas}, \bibinfo{person}{Jamie Hall}, \bibinfo{person}{Noam Shazeer},
  \bibinfo{person}{Apoorv Kulshreshtha}, \bibinfo{person}{Heng-Tze Cheng},
  \bibinfo{person}{Alicia Jin}, \bibinfo{person}{Taylor Bos},
  \bibinfo{person}{Leslie Baker}, \bibinfo{person}{Yu Du}, {et~al\mbox{.}}}
  \bibinfo{year}{2022}\natexlab{}.
\newblock \showarticletitle{Lamda: Language models for dialog applications}.
\newblock \bibinfo{journal}{\emph{arXiv preprint arXiv:2201.08239}}
  (\bibinfo{year}{2022}).
\newblock


\bibitem[Touvron et~al\mbox{.}(2023)]%
        {touvron2023llama}
\bibfield{author}{\bibinfo{person}{Hugo Touvron}, \bibinfo{person}{Thibaut
  Lavril}, \bibinfo{person}{Gautier Izacard}, \bibinfo{person}{Xavier
  Martinet}, \bibinfo{person}{Marie-Anne Lachaux},
  \bibinfo{person}{Timoth{\'e}e Lacroix}, \bibinfo{person}{Baptiste
  Rozi{\`e}re}, \bibinfo{person}{Naman Goyal}, \bibinfo{person}{Eric Hambro},
  \bibinfo{person}{Faisal Azhar}, {et~al\mbox{.}}}
  \bibinfo{year}{2023}\natexlab{}.
\newblock \showarticletitle{Llama: Open and efficient foundation language
  models}.
\newblock \bibinfo{journal}{\emph{arXiv preprint arXiv:2302.13971}}
  (\bibinfo{year}{2023}).
\newblock


\bibitem[Vaswani(2017)]%
        {vaswani2017attention}
\bibfield{author}{\bibinfo{person}{A Vaswani}.}
  \bibinfo{year}{2017}\natexlab{}.
\newblock \showarticletitle{Attention is all you need}.
\newblock \bibinfo{journal}{\emph{Advances in Neural Information Processing
  Systems}} (\bibinfo{year}{2017}).
\newblock


\bibitem[Yildiz et~al\mbox{.}(2022)]%
        {yildiz2022decaf}
\bibfield{author}{\bibinfo{person}{Orcun Yildiz}, \bibinfo{person}{Matthieu
  Dreher}, {and} \bibinfo{person}{Tom Peterka}.}
  \bibinfo{year}{2022}\natexlab{}.
\newblock \showarticletitle{Decaf: Decoupled Dataflows for In Situ Workflows}.
\newblock In \bibinfo{booktitle}{\emph{In Situ Visualization for Computational
  Science}}. \bibinfo{publisher}{Springer}, \bibinfo{pages}{137--158}.
\newblock


\bibitem[Yildiz et~al\mbox{.}(2024a)]%
        {yildiz2024extreme}
\bibfield{author}{\bibinfo{person}{Orcun Yildiz}, \bibinfo{person}{Amal
  Gueroudji}, \bibinfo{person}{Julien Bigot}, \bibinfo{person}{Bruno Raffin},
  \bibinfo{person}{Rosa~M Badia}, {and} \bibinfo{person}{Tom Peterka}.}
  \bibinfo{year}{2024}\natexlab{a}.
\newblock \showarticletitle{Extreme-scale workflows: A perspective from the
  JLESC international community}.
\newblock \bibinfo{journal}{\emph{Future Generation Computer Systems}}
  \bibinfo{volume}{161} (\bibinfo{year}{2024}), \bibinfo{pages}{502--513}.
\newblock


\bibitem[Yildiz et~al\mbox{.}(2024b)]%
        {yildiz2024wilkins}
\bibfield{author}{\bibinfo{person}{Orcun Yildiz}, \bibinfo{person}{Dmitriy
  Morozov}, \bibinfo{person}{Arnur Nigmetov}, \bibinfo{person}{Bogdan Nicolae},
  {and} \bibinfo{person}{Tom Peterka}.} \bibinfo{year}{2024}\natexlab{b}.
\newblock \showarticletitle{Wilkins: HPC in situ workflows made easy}.
\newblock \bibinfo{journal}{\emph{Frontiers in High Performance Computing}}
  \bibinfo{volume}{2} (\bibinfo{year}{2024}), \bibinfo{pages}{1472719}.
\newblock


\bibitem[Yu and Buyya(2005)]%
        {yu2005taxonomy}
\bibfield{author}{\bibinfo{person}{Jia Yu} {and} \bibinfo{person}{Rajkumar
  Buyya}.} \bibinfo{year}{2005}\natexlab{}.
\newblock \showarticletitle{A taxonomy of workflow management systems for grid
  computing}.
\newblock \bibinfo{journal}{\emph{Journal of grid computing}}
  \bibinfo{volume}{3} (\bibinfo{year}{2005}), \bibinfo{pages}{171--200}.
\newblock


\end{thebibliography}










\end{document}